\documentclass[pre,aps,floats,superscriptaddress,floatfix]{revtex4}
\usepackage{amssymb,amsmath}
\usepackage{amsmath,amssymb}
\usepackage{graphicx}
\usepackage{psfrag}

\def\beq{\begin{equation}}
\def\eeq{\end{equation}}
\def\bea{\begin{eqnarray}}
\def\eea{\end{eqnarray}}
\begin{document}
\title{Active to absorbing state phase transition in the presence of a
fluctuating environment: Feedback and universality}
\author{Niladri Sarkar}\email{niladri.sarkar@saha.ac.in}
\author{Abhik Basu}\email{abhik.basu@saha.ac.in}
\affiliation{Condensed Matter Physics Division, Saha Institute of
Nuclear Physics, Calcutta 700064, India}

\date{\today}
\begin{abstract}
We construct and analyse a simple {\em reduced} model to study the
effects of the interplay between a density undergoing an
active-to-absorbing state phase transition (AAPT) and a fluctuating
environment in the form of a broken symmetry mode coupled to the
density field in any arbitrary dimension. We show, by using
perturbative renormalisation group calculations, that {\em both} the
effects of the environment on the density and the latter's feedback
on the environment influence the ensuing universal scaling behaviour
of the AAPT at its extinction transition. Phenomenological
implications of our results in the context of more realistic natural
examples are discussed.
\end{abstract}

\maketitle

\section{Introduction}

The phenomena of active to absorbing state phase transition (AAPT)
forms a paradigmatic example of non-equilibrium phase transitions.
The enumeration of the scaling exponents that characterise the AAPT
and the corresponding universality classes are topics of intense
research activities at present~\cite{uwe-hans-review}. It is now
generally believed, as enunciated in what is known as the {\em
Directed Percolation Hypothesis}~\cite{dphp},  that in the absence
of any special symmetry, conservation law, quenched disorder or
long-ranged interactions the AAPT belongs to the directed
percolation (DP) universality class, as long as there is a single
absorbing state. Well-known examples of models belonging to the DP
universality class include the Gribov~\cite{gribov} process or the
epidemic process with recovery and the stochastic formulations of
the predator prey automaton models~\cite{uwe-hans-review}. Continuum
versions of models belonging to the DP universality class are
described formally by the Reggeon field
theory~\cite{rft1,rft2,rft3}, which is a stochastic multiparticle
process that describes the essential features of local growth
processes of populations in a uniform environment near their
extinction threshold \cite{popdyn2,popdyn3}. The model parameters of
the Reggeon field theory depend on the embedding environment and are
chosen as constants; thus the environment is considered uniform and
its fluctuations are ignored there.

The DP hypothesis and the associated DP universality class are
believed to be very general and robust. Nonetheless, it is
reasonable to expect that environmental fluctuations should affect
the universal scaling properties of the AAPT in the DP universality
class. For instance, the critical scaling behaviour of a density
$\phi$ undergoing an AAPT in the presence of fluctuating
environments  has been shown in \cite{nil1}; see
also~\cite{anton1,anton2} for related studies. In \cite{nil1}
different models were used to describe the fluctuating environments
namely (i) the randomly stirred fluid modeled by the Navier-Stokes
equation and (ii) fluctuating surface modelled by the
Kardar-Parisi-Zhang equation or (iii) the Edward-Wilkinson equation.
In all these cases the dynamic exponent of the environment was found
to be either same as the DP dynamic exponent (strong dynamic
scaling) or different from that of the dynamic exponent of the
percolating field (weak dynamic scaling) resulting in non DP
behaviour.  Not surprisingly, critical exponents belonging to new
universality classes were found. From a technical perspective, in
all these examples (a) the environment is modeled by a long-ranged
noise driven conserved hydrodynamic variable, e.g., a velocity field
or a fluctuating surface (equivalently a Burgers velocity field),
and (b) in all the cases the feedback of the density field
undergoing AAPT on the environment is ignored, i.e., the dynamics of
the environment is assumed to be {\em autonomous}. Both of these
features are certainly special cases, since the origin of
environmental fluctuations may be very different from a fluctuating
conserved variable (e.g., a Navier-Stokes velocity field). For
instance, the environment may contain {\em broken symmetry fields},
e.g., elastic deformations of crystals or liquid crystals,
fluctuations in membranes, deformations in an ordered suspension of
orientable particles etc. Secondly, the dynamics of the environment,
in general, is {\em not} expected to be autonomous; instead it
should be affected by the density $\phi$ that undergoes an AAPT.
Since $\phi$ is expected to have a long-ranged correlation near the
AAPT transition, it will {\em effectively} act as an {\em additional
stochastic noise source} with long-ranged correlation, which may
alter the scaling properties of the environmental dynamics. These
issues are likely to be important in some recent experiments on
living cells~\cite{melano}, discussing two possible
symmetry-determined orientationally ordered states: (a) the {\em
active vectorial} or {\em polar} order, where the (elongated) cells
are oriented along a mean direction $\bf\hat p$ with $\bf \hat p$
and $-\bf\hat p$ being {\em inequivalent},  and (b) {\em active
apolar} or {\em nematic} order where $\bf\hat p$ and $-\bf\hat p$
are equivalent. At the continuum mesoscopic level, these systems are
described typically by coupled dynamical equations
 of the particle density and local orientational order parameter (and
also a hydrodynamic velocity if the system is momentum conserving);
see, e.g., see Ref.~\cite{review} for recent reviews and detailed
discussions concerning these systems.  A similar example is the
orientational order of the magnetotactic bacteria along the earth's
magnetic field lines~\cite{magneto}. In such systems, if the
experimental time scales are much larger than the natural birth
(reproduction/cell division) and death time-scales, then such
nonconservation processes are likely to affect the emerging
macroscopic properties.

Motivated by the  theoretical issues of the effects of feedback to
the fluctuating environment on the universal scaling properties of
AAPT, in this paper we propose and study a simple reduced model for
AAPT in the presence of an environment modeled by a broken symmetry
field described by a vector field $\bf v$, whose dynamics in turn is
affected by $\phi$ (feedback). Thus, this study is substantially
different from and complementary to Ref.~\cite{nil1} in having an
environment dynamics that is no longer autonomous due to the
feedback, a situation not considered in Ref.~\cite{nil1}. Apart from
the consideration of the feedback of $\phi$ on the dynamics of the
environment, we point out a crucial technical difference that the
environmental dynamics has a {\em short-ranged} Gaussian noise (see
below), unlike in Ref.~\cite{nil1}, where the corresponding noises
are all considered to be spatially long-ranged. Our principal result
here is that the scaling behaviour of the system near the extinction
transition of the AAPT is, in general, affected by the feedback on
the environment and hence non-DP like. In general, depending upon
the location of the system in the phase space, one may encounter
{\em strong dynamic scaling} (when both $\phi$ and $\bf v$ have the
same dynamic exponents relating spatial and temporal scalings) or
{\em weak dynamic scaling}, when the two dynamic exponents are
different. Our results here should help us understand the general
effects of mutual dynamical interaction between a density field and
the embedding environment for more realistic but complicated
situations. The rest of the paper is organised as follows: In
Sec.~\ref{model}, we set up our model following a brief review the
DP universality class. Then we do a detailed dynamic renormalisation
group (DRG) analysis of our model to obtain the scaling exponents at
the AAPT in Sec.~\ref{scale}. Finally, in Sec.~\ref{summ} we
conclude and summarise our results.

\section{Dynamical Model}
\label{model}

In order to address the issues as mentioned above systematically we
construct a simple model in which a density field $\phi$ undergoing
AAPT is coupled to a fluctuating broken symmetry field, represented
by a vector field $\bf v$, which acts as the environment. A feedback
from the density $\phi$ to the dynamics of $\bf v$ is the
distinguishing feature of the present model. Before we discuss it in
details, in order to set up the background, we briefly review the
problem of extinction transition of a single species in a uniform
environment and the scaling exponents at the corresponding AAPT as
described by the DP universality class or Gribov process.

\subsection{Directed Percolation model}\label{review}

Let us consider a population dynamics with a population growth rate
depending linearly on the local species density  and a death rate
controlled by the square of the local density (qualitatively representing
death due to overcrowding) undergoing a non-equilibrium active to
absorbing state (i.e., species extinction) phase transition whose
long distance large time properties are well-described by the DP
universality class. In terms of a local particle density $\phi({\bf
x},t)$, the Langevin equation that describes such a population
dynamics is given by [see, e.g., Ref.~\cite{uwe-hans-review}]
\begin{equation}
\frac{\partial\phi}{\partial t} = D\nabla^2\phi + \lambda_g\phi -
\lambda_d\phi^2 + \sqrt\phi \zeta,\label{dpeq}
\end{equation}
where diffusive modes of the density is included with $D$ as the
diffusion coefficient, $\lambda_g$ is the growth rate and
$\lambda_d$ the decay rate. Stochastic function $\zeta({\bf x},t)$
is a zero-mean, Gaussian distributed white noise with a variance
\begin{equation}
\langle\zeta({\bf x},t)\zeta(0,0)\rangle = 2D_2 \delta ({\bf
x})\delta (t). \label{vari1}
\end{equation}
The in-principle existence of an absorbing state ($\phi=0$) in the
system is ensured by the multiplicative nature of the effective
noise. We may extract the characteristic length $\xi \sim \sqrt
{D/|\lambda_g|}$ and diffusive time scale $t_c\sim \xi^2/D\sim
1/|\lambda_g|$ on dimensional ground, from Eq.~(\ref{dpeq}), both of
which diverge upon approaching the critical point at $\lambda_g=0$.
 We
then define the critical exponents in the usual way
\cite{uwe-hans-review}
\begin{equation}
\langle \phi ({\bf x},t\rightarrow\infty)\rangle \sim
\lambda_g^\beta,\;\;\langle \phi({\bf x},t)\rangle \sim
t^{-\alpha}\; (\lambda_g=0),\;\; \xi\sim \lambda_g^{-\nu},\;\;
t_c\sim \xi^z_\phi/D\sim \lambda_g^{-z_\phi\nu}, \label{scaling}
\end{equation}
yielding the mean-field values for the scaling exponents
\begin{equation}
\beta=1,\alpha=1,\nu=1/2,\,\, {\mbox {and}},\,\, z_\phi=2.\label{mean}
\end{equation}
In addition, the anomalous dimension $\eta$, which characterises the
spatial scaling of the two-point correlation function, is zero
\cite{uwe-hans-review}. Whether or not fluctuations change the
scaling behaviour of the DP problem from their mean-field values,
characterised by (\ref{mean}) is an important question here.  The DP
problem, as modeled by Janssen-de Dominicis action functional
corresponding to the Langevin Eq.~(\ref{dpeq}), is invariant under
the rapidity symmetry given by $\hat\phi ({\bf x},t) \leftrightarrow
\phi({\bf x},-t)$~\cite{uwe-hans-review}, where $\hat\phi$ is the
dynamic conjugate variable~\cite{uwe-hans-review}; see below also.
This invariance formally defines the DP universality class. All
models belonging to the DP universality class are invariant under
the rapidity symmetry asymptotically. In order to account for the
fluctuation effects, which are expected to affect the mean-field
exponent values (\ref{mean}), DRG calculations have been performed
over an equivalent path integral description of the Langevin
Eq.~(\ref{dpeq})~\cite{uwe-hans-review}. By using one-loop
renormalised theory with a systematic $\epsilon$-expansion,
$\epsilon=d_c-d$, where the upper critical dimension $d_c=4$ for
this model, one obtains \cite{uwe-hans-review},
\begin{equation}
z=2-\epsilon/12,\eta=\epsilon/12 \,\,\mbox{and}\,\, {1 \over \nu}=2
+ \epsilon/4. \label{dpexpo}
\end{equation}
The set of exponents (\ref{dpexpo}) formally constitute and
characterise the DP universality class.
 Recent studies suggest that the DP universality class
is fairly robust, a feature formally known as the directed
percolation (DP) hypothesis~\cite{dphp}.  Only when one or more
conditions of the DP hypothesis are violated, one finds new
universal properties. For instance, the presence of long range
interactions are known to modify the scaling behaviour:
Ref.~\cite{field2} examines the competition between short and long
ranged interactions, and identified four different possible phases.
Subsequently, Refs.~\cite{anton1,anton2,nil1} have shown how
fluctuating environments driven by spatially long-ranged noises (but
with autonomous dynamics) may modify the scaling behaviour of the DP
universality. In the present work, we extend and complement the
existing results by considering a model study (without any
long-ranged noise) that considers the effects of feedback of the
species density on the environment dynamics. It is expected that
such additional couplings between the species density and the
environment may alter the universal behaviour at the AAPT. Our
perturbative results below confirm this.

%\subsection{Broken symmetry} \label{rand1}

\subsection{Extinction transition in the presence of a broken symmetric
field} \label{rand}

Having set up the background of our work here, in this subsection we
set up the equations of motion for our model of the the density
field $\phi({\bf x},t)$ undergoing AAPT coupled with the broken
symmetry field ${\bf v} ({\bf x},t)$ in the hydrodynamic limit,
retaining minimal but relevant coupling terms connecting the
dynamics of $\phi$ and $\bf v$. A broken symmetry field, also known
as a {\em Goldstone variable} in the literature is a deformation of
an ordered state that originates in a system due to the breakdown of
a continuous symmetry. Well-known examples of broken symmetry states
include crystals (broken translational invariance), nematic liquid
crystals (broken rotational invariance), Heisenberg ferromagnetic
systems (broken rotational invariance in the order parameter
space)~\cite{chaikin}. A broken symmetry mode necessarily has a
life-time of a fluctuation that diverges in the zero wavevector
limit, reflecting the simple fact that cost of configuration energy
associated with the creation of a broken symmetry mode with a given
wavelength vanishes as the wavelength of the fluctuation diverges.
There are, however, no conservation law associated with a broken
symmetry variable. A broken symmetry variable may have a variety of
symmetry, depending upon the actual physical system concerned. For
instance, the local displacement fields~\cite{chaikin,landau}, the
broken symmetry variables in a crystal are invariant under shifts by
constant amounts, whereas, the Frank director field, which are the
relevant broken symmetry variables in a nematic liquid crystal, are
invariant under a combined rotation of the coordinate system and the
director fields.
 From a general theoretical point of view, it is interesting
to study the universal behaviour of AAPT in contact with a broken
symmetry mode. Apart from this, studies on AAPT in contact with
broken symmetries are potentially relevant in exploring the
universal properties of the extinction transitions in a bacteria
colony populated by bacteria in their orientated states, e.g., polar
or nematic. The time-evolution of the polar or nematic order
parameter should be generically coupled to the density undergoing
AAPT, and hence may affect the scaling at the AAPT. Since the order
parameter field is a broken symmetry field, its correlation function
is scale invariant, displaying universal scaling. Whether the scale
invariant density field at the AAPT modifies the scaling of the
order parameter fields through the mutual dynamical couplings is an
associated relevant question. While we are motivated by these
examples, in the present work, we do not intend to model a specific
case of broken symmetry variable as the environment; rather, it may
be considered as a toy model for AAPT in contact with a broken
symmetry variable with a simple structure. To this effect, we
enforce a simple invariance on the broken symmetry variable $\bf v$
in the model by demanding invariance under ${\bf v}\rightarrow {\bf
v} + {\bf v}_0$~\cite{crystal}. Thus any coupling between $\phi$ and
$\bf v$ should involve ${\boldsymbol \nabla}\cdot {\bf v}$. Such
considerations allow us to write down the dynamical equation for
$\phi$: This is essentially same as Eq.~(\ref{dpeq}), supplemented
by a  symmetry-allowed coupling term involving ${\boldsymbol
\nabla}\cdot {\bf v}$ and $\phi$. The resulting equation of motion
for $\phi$ up to the lowest order in spatial gradients takes the
form
 \bea {\partial \phi \over \partial
t}=\lambda_g \phi -\lambda_d\phi^2 +D\nabla^2\phi +
\lambda_1\phi{\boldsymbol \nabla}\cdot {\bf v} + \sqrt{\phi}\xi,
\label{dpsb} \eea where $\lambda_1$ is the coefficient describing
the most dominant lowest order coupling that couples a vector field
$\bf v$ with a scalar field $\phi$, $\lambda_g$ and $\lambda_d$ are
the growth and decay coefficients of the density $\phi$ and $\xi$ is
the gaussian distributed white noise with a variance as given by
Eq.~(\ref{vari1}).

To complete the dynamical description of our model, we now need a
corresponding equation for $\bf v$. We use a simple relaxational
dynamics for $\bf v$. To obtain the appropriate dynamical equation,
we start with a free energy functional: We assume that the energy
associated with the configurations of $\bf v$ are given by
 \bea
\mathcal{F}={1 \over 2}\int d^dx
[\lambda(\nabla_iv_j)^2+2\chi({\boldsymbol \nabla}\cdot {\bf v})\phi],
\label{free},
 \eea
  where $\lambda>0$ is the
stiffness modulus (akin to the elastic modulii for a crystal or the
Frank elastic constants for nematic liquid crystals)  and $\chi$ is
the coupling constant for the bilinear coupling between $\bf v$ and
$\phi$. Assuming a non-conserved relaxational dynamics (model A in
the language of Ref.~\cite{halp}) the stochastically driven
Langevin equation for ${\bf v}$ becomes ${\partial v_i \over
\partial t}=-\hat\Gamma{\delta \mathcal{F} \over \delta v_i}+f_i$, where $f_i$ is a zero-mean Gaussian noise,
$\hat\Gamma$ is a kinetic coefficient (set to unity below). With the
choice of $\mathcal F$ as above, we find \bea \frac{\partial
v_i}{\partial t}&=& \lambda \nabla^2 v_i + \chi \nabla_i\phi + f_i.
\label{equ}\eea For systems in equilibrium, the variance of $f_i$
would have been related to it through the
Fluctuation-Dissipation-Theorem (FDT)~\cite{chaikin}. However, the
present system being out-of-equilibrium, where there is no FDT, the
variance of ${\bf f}$ is unrelated to $\hat\Gamma$. We choose \bea
\langle f_i({\bf x},t)f_j(0,0)\rangle =
2D_0\delta^d(x)\delta(t)\delta_{ij}. \label{fnoise}\eea From
symmetry point of view, clearly, our model Eqs.~(\ref{dpsb}) and
(\ref{equ}) are not invariant under ${\bf v\rightarrow -\bf v}$.
Thus, this is reminiscent of polar (or vectorial) symmetry of
Ref.~\cite{toner}.

The vector field $v_i$ being a broken symmetry field has a dynamics
that is generically scale invariant, characterised by a set of
scaling exponents. They are defined via the correlation function
 \bea
 \langle v_i ({\bf x},t)v_j(0,0)\rangle = |{\bf x}|^{2-d-\eta_v}
 \psi_{ij}^v (|{\bf x}|^{z_v}/t),
 \eea
 where $\eta_v$ and $z_v$ are the anomalous dimension and dynamic
 exponent, respectively of $\bf v$, and $\psi^v_{ij}$ is a dimensionless
 scaling function of its argument. Ignoring the coupling with
 $\phi$, exponents $\eta_v=0$ and $z_v=2$ are known exactly. Whether
 the coupling with $\phi$ alters these exponents is a question that
 we address here within a one loop perturbative calculation.

  Redefining coefficient $\lambda_g=D\tau$
and $\lambda_d={Dg_2 \over 2}$ for calculational convenience, Eq. (\ref{dpsb}) may be written
as
 \bea {\partial\phi \over \partial
t}=D(\tau +\nabla^2)\phi - {Dg_2 \over 2}\phi^2 +
\lambda_1({\boldsymbol \nabla}\cdot {\bf v})\phi +\sqrt{\phi}\xi,
\label{dpsb2}
 \eea
  which redefines the
critical point as renormalised $\tau=0$. In the mean field picture
(dropping all nonlinearities) at $\tau=0$, density $\phi$ undergoes
an AAPT displaying the mean-field DP universal behaviour with
critical exponents given by Eq.~(\ref{mean}) above. Whether or not
the nonlinear coupling terms $\lambda_1$ and $Dg_2$, together with
the (linear) feedback term with coefficient $\chi\hat\Gamma$ are
able to alter the mean-field universal behaviour can only be
answered by solving the full coupled equations (\ref{dpsb2}) and
(\ref{equ}). Their overall nonlinear nature rules out the
possibility of any exact solution. A well-established framework for
addressing this issue systematically is the standard implementation
of DRG procedure, based on a one-loop perturbative expansion in the
coupling constants $\lambda_1$ and $Dg_2$ about the linear theory.
The resulting perturbative corrections of the different (bare) model
parameters may then be used to construct the renormalised
correlation functions.

We begin with the Janssen-De Dominics generating
functional~\cite{janssen} corresponding to the Langevin
Eqs.~(\ref{equ}) and (\ref{dpsb2}) and the noise variances
(\ref{vari1}) and (\ref{fnoise}), which allows us to describe the
dynamics as a path integral over the relevant dynamical fields in
the system. For the convenience of calculations that follow, we
redefine $i\hat v_i\rightarrow\hat v_i, i\hat\phi\rightarrow
\beta_1\hat\phi$ and $\phi\rightarrow \beta_2\phi$,
$\beta_1^2D_2\beta_2= {Dg_1 \over 2}$, $\beta_1\beta_2=1,
\lambda_g=\tau D$ and $\lambda_d\beta_2={Dg_2 \over
2}$~\cite{dpcalc}. Writing the generating functional as
 \bea \langle\mathcal{Z}\rangle_f=\int D\phi D\hat\phi DvD\hat v \exp[-{\mathcal S}], \label{gen} \eea
  where ${\mathcal S}$ is the action functional of the system. The expression for ${\mathcal S}$ can
be written as
 \bea {\mathcal S}
&=& -{Dg_1 \over 2}\int {d^dk \over
(2\pi)^d}\frac{d^dq}{(2\pi)^d}\int {d\omega \over
2\pi}\frac{d\Omega}{2\pi}\hat\phi_{{-\bf k},-\omega}
\hat\phi_{{\bf q},\Omega} \phi_{{\bf k-q},\omega-\Omega} + {Dg_2
\over 2}\int {d^dk \over (2\pi)^d}\frac{d^dq}{(2\pi)^d}\int
{d\omega \over 2\pi}\frac{d\Omega}{2\pi}\hat\phi_{{\bf -k},-\omega}
\phi_{{\bf q},\Omega}\phi_{{\bf k-q},\omega-\Omega}
\nonumber \\
&& + \int {d^dk \over (2\pi)^d} \int {d\omega \over
2\pi}\hat\phi_{{\bf -k},-\omega} \left\{i\omega\phi_{{\bf k},\omega}
+ D(-\tau +k^2)\phi_{{\bf k},\omega} -
i\lambda_1\int\frac{d^dq}{(2\pi)^d}\frac{d\Omega}{2\pi}\hat\phi_{-{\bf k},-\omega}
q_lv_l({\bf q},\Omega)\phi_{{\bf k}-{\bf q},\omega-\Omega}\right\} \nonumber \\
&& -\int {d^dk \over (2\pi)^d} \int {d\omega \over 2\pi}D_0\hat
v_i(-{\bf k},-\omega) \hat v_i({\bf k},\omega)+ \int {d^dk \over
(2\pi)^d} \int {d\omega \over 2\pi}\hat v_i({\bf -k},-\omega)
\left\{(i\omega +\lambda k^2)v_i({\bf k},\omega) -i\chi
k_i\phi_{{\bf k},\omega}\right\}. \label{act} \eea Here ${\bf k},
{\bf q}$ represent momenta and $\omega, \Omega$ represent
frequencies in the Fourier space. The first two terms in
Eq.~(\ref{act}) have different coefficients ${Dg_1 \over 2}$ and
${Dg_2 \over 2}$ which shows the breakdown of invariance under
rapidity symmetry~\cite{uwe-hans-review} as a result of the
couplings $\lambda_1$ and $\chi$. Notice that by rescaling time
we may absorb the coefficient $\lambda$. This explains the lack of
renormaliization for $\lambda$ (see below for details).

 Before we embark upon the
detailed calculation, let us note the following: First of all, the
action functional (\ref{act}) is no longer invariant under the
rapidity symmetry; the coupling with the broken symmetry field $\bf
v$ explicitly breaks it. Given our wisdom from equilibrium critical
phenomena and equilibrium critical dynamics, new universal behaviour
is expected, provided the dynamical couplings  between $v_i$ and
$\phi$ are relevant. As a result, scaling exponents should have
values different from their values in the DP universality class
given by (\ref{dpexpo}). When the coupling $\chi$ [the feedback term
in Eq.~(\ref{equ})] is zero, the dynamics of $\bf v$ becomes
autonomous, i.e., independent of $\phi$. Evidently, in this case,
$z_v=2$, where as $z_\phi$ may or may not be 2. Thus, one may
encounter both {\em weak} and {\em strong} dynamic scaling. On the
other hand, when (renormalised) $\chi\neq 0$, the dynamics of $\bf
v$ is no longer autonomous; it gets affected by the dynamics of
$\phi$, such that $z_v$ may be different from 2. Whether or not
$z_\phi$ is same as $z_v$ can be ascertained only after a detailed
calculation that we present below.

To start with we note that the roles of the ({\em bare} or {\em
unrenormalised}) coupling constants in an ordinary perturbative
expansion of the present model are played by $u=g_1g_2$ and
$w={\lambda_1^2 \over D^3}$.  Our model has upper critical dimension
$d_c=4$, such that both the coupling constants $u$ and $w$ become
dimensionless at $d=4$, and the mean-field exponents (\ref{mean})
are to provide quantitatively correct description of scaling for
$d\geq 4$. We set up a renormalised perturbative expansion in
$\epsilon = 4-d$ up to the one-loop order. To ensure ultra-violet
(UV) renormalisation of the present model, we render finite all the
non-vanishing two-, three-point vertex functions by introducing
multiplicative renormalisation constants. This procedure is standard
and well-documented in the literature, see, e.g., Ref.~\cite{drg}.
Here, the vertex functions of different orders are formally defined
by appropriate functional derivatives of the vertex generating
functional $\Gamma[\phi,\hat\phi,v_i,\hat v_i]$ which is the
Legendre transformation of $\log {\mathcal Z}$~\cite{drg}.  The bare
values of the different vertex functions can be easily read off the
action functional (\ref{act}) and are given by (after separating out
the various $\delta$-functions associated with spatial and temporal
translation invariance)
 \bea
 &&\frac{\delta^2 \Gamma}{\delta\phi ({\bf k},\omega)\delta\hat\phi({\bf -k},-\omega)}=
 \Gamma_{\phi\hat\phi}=i\omega + D(-\tau + k^2),\\
 &&\frac{\delta^2 \Gamma}{\delta v_i ({\bf k},\omega)\delta\hat v_j({\bf -k},-\omega)}=
 \Gamma_{v_i\hat v_j}=(i\omega +\lambda k^2)\delta_{ij}, \\
 &&\frac{\delta^2 \Gamma}{\delta \hat v_i ({\bf k},\omega)\delta \hat v_j({\bf -k},-\omega)}=
 \Gamma_{\hat v_i\hat v_j}= 2D_0\delta_{ij}, \\
 &&\frac{\delta^2 \Gamma}{\delta \hat v_i ({\bf -k},-\omega)\delta\phi({\bf k},\omega)}=
 \Gamma_{\hat v_i\phi}=-i\chi k_i,\\
&&\frac{\delta^3 \Gamma}{\delta \hat\phi({\bf q}_1,\omega_1)
\delta\hat\phi ({\bf q}_2,\omega_2)\delta\phi({\bf -q}_1-{\bf
q}_2,-\omega_1 -\omega_2)}=\Gamma_{\hat
\phi\hat\phi\phi}=-\frac{Dg_1}{2},\\
&&\frac{\delta^3 \Gamma}{\delta \hat\phi({\bf q}_1,\omega_1)
\delta\phi ({\bf q}_2,\omega_2)\delta\phi({\bf -q}_1-{\bf
q}_2,-\omega_1 -\omega_2)}=\Gamma_{\hat
\phi\phi\phi}=\frac{Dg_2}{2},\\
&&\frac{\delta^3 \Gamma}{\delta v_i ({\bf k},\omega)\delta\hat\phi
({\bf q},\Omega)\delta \phi ({\bf -k
-q},-\omega-\Omega)}=\Gamma_{v_i\hat\phi\phi}=-i\lambda_1 k_i.
 \eea

%{\bf [remaining 3 and 4 pt vertex functions?]}

\section{renormalisation group calculations and the scaling exponents}
\label{scale}

In order to renormalise the vertex functions by carrying out the one
loop integrals we choose $\tau=\mu^2$ as our appropriate
normalization point, where $\mu$ is an intrinsic momentum scale of
the renormalised theory. This will allow us to find the scale
dependence of the renormalised correlation or vertex functions on
$\mu$ by using the multiplicative $Z$-factors for the fields and the
parameters in the model. These $Z$-factors are useful in absorbing
all the ultraviolet divergences arising from the one loop
diagrammatic corrections thus giving us an effective finite theory.
Formally, the $Z$-factors present in this model are defined as \bea
&&\phi=Z_\phi \phi_R \,,\, v=Z_v v_R\,,\,\hat v=Z_{\hat v}\hat
v_R\,,\,\hat\phi=Z_{\hat\phi}\hat\phi_R\,,\,D=Z_DD_R\,,\,
\lambda_1=Z_{\lambda_1}\lambda_{1R} \,,\, g_1=Z_{g_1}g_{1R}\,,\,
g_2=Z_{g_2}g_{2R}\,,\,\tau=Z_\tau \tau_R,\nonumber \\
&&\lambda = Z_\lambda \lambda_R,\chi=Z_\chi\chi_R, \label{zfacdef}
 \eea
 where a subscript $R$ refers to a renormalised quantity.
The different $Z$-factors may be enumerated from the following
conditions on the renormalised vertex functions:
 \bea
 \frac{\partial\Gamma_{\hat\phi\phi}}{\partial\omega}|_{({\bf
 k} =0,\omega=0)}&=&i,\\
 \frac{\partial\Gamma_{\hat\phi\phi}}{\partial k^2}|_{({\bf
 k}=0,\omega=0)} &=& D_R,\\
 \Gamma_{\hat\phi\phi}({\bf k}=0,\omega=0)&=&D_R\tau_R,\\
 \frac{\partial\Gamma_{\hat v_i v_j}}{\partial\omega}|_{({\bf
 k} =0,\omega=0)}&=&i\delta_{ij},\\
\frac{\partial\Gamma_{\hat v_i v_j}}{\partial k^2}|_{({\bf
 k}=0,\omega=0)} &=& \lambda\delta_{ij},\\
 \Gamma_{\hat\phi\hat\phi\phi}({\bf k}=0,{\bf
 q}=0,\omega=0,\Omega=0)&=& -\frac{D_Rg_{1R}}{2},\\
 \Gamma_{\hat\phi\phi\phi}({\bf k}=0,{\bf
 q}=0,\omega=0,\Omega=0)&=& \frac{D_Rg_{2R}}{2},\\
\frac{\partial}{\partial k_i} \Gamma_{v_i\hat\phi\phi}|_{({\bf
k}=0,{\bf
 q}=0,\omega=0,\Omega=0)}&=& -i\lambda_{1R},\\
\Gamma_{\hat v_i\hat v_j} ({\bf k=0},\omega=0)&=& -2D_0\delta_{ij}.
 \label{renorm}
 \eea
%{\bf other vertex functions?}

% {\bf needs correction: The last two in Eqs.~(\ref{renorm}) shows that $\lambda$ is not an
 %independent parameter in the theory and it can be assumed to remain unrenormalised;
 %we thus set $Z_\lambda=1$.}

There are 11 $Z$-factors defined in Eq.~(\ref{zfacdef}) above, as
compared to the 9 renormalisation conditions on the renormalised
vertex functions, as given in Eq.~(\ref{renorm}). Thus, two of the
$Z$-facors defined above are redundant. Without any loss of
generality, we set $Z_\phi=Z_{\hat\phi}$ and $Z_v=Z_{\hat v}$.
 Explicit forms for the $Z$-factor are given by
\bea
Z_\phi &=&
Z_{\hat\phi}= 1 + {g_1g_2\mu^{-\epsilon} \over 8\epsilon}\frac{1}{16\pi^2} -
{D_0\lambda_1^2\mu^{-\epsilon} \over \lambda(\lambda +D)^2\epsilon}\frac{1}{16\pi^2}
-{\lambda_1\chi g_1(3D+\lambda)\mu^{-\epsilon} \over 4D(D+\lambda)^2\epsilon}\frac{1}{16\pi^2},
\nonumber \\
Z_D &=& 1-{g_1g_2\mu^{-\epsilon} \over 8\epsilon}\frac{1}{16\pi^2} + {\lambda_1\chi g_1
(7D^2+4\lambda D+\lambda^2)\mu^{-\epsilon} \over 4D(D+\lambda)^3\epsilon}\frac{1}{16\pi^2}
+ {2DD_0\lambda_1^2\mu^{-\epsilon} \over \lambda(D+\lambda)^3\epsilon}\frac{1}{16\pi^2}, \nonumber \\
Z_\tau &=& 1+{3g_1g_2\mu^{-\epsilon} \over 8\epsilon}\frac{1}{16\pi^2} -
{2DD_0\lambda_1^2\mu^{-\epsilon} \over \lambda(D+\lambda)^3}\frac{1}{16\pi^2}
-{\lambda_1\chi g_1(5D+3\lambda)\mu^{-\epsilon} \over 2D(D+\lambda)^2\epsilon}
\frac{1}{16\pi^2}-{\lambda_1g_1\chi (7D^2+4D\lambda +\lambda^2)\mu^{-\epsilon}
\over 4D(D+\lambda)^3\epsilon}\frac{1}{16\pi^2}, \nonumber \\
Z_{g_2} &=& 1+{3g_1g_2\mu^{-\epsilon} \over 4\epsilon}\frac{1}{16\pi^2}-{D_0\lambda_1^2(4\lambda
+5D)\mu^{-\epsilon} \over \lambda D(\lambda +D)^2\epsilon}\frac{1}{16\pi^2}-
{2DD_0\lambda_1^2\mu^{-\epsilon} \over \lambda(\lambda +D)^3\epsilon}\frac{1}{16\pi^2}
+{\lambda_1^2g_1\chi^2(2D+\lambda)\mu^{-\epsilon} \over \lambda D^2g_2
(\lambda +D)^2\epsilon}\frac{1}{16\pi^2} \nonumber \\
&&+{\lambda_1g_1\chi(3D+\lambda)\mu^{-\epsilon}
\over D(D+\lambda)^2\epsilon}\frac{1}{16\pi^2}
-{4\lambda_1^3\chi D_0\mu^{-\epsilon}
\over \lambda^2D^2g_2(D+\lambda)\epsilon}\frac{1}{16\pi^2}+{\lambda_1\chi g_1
(5D^2+12\lambda D+5\lambda^2)\mu^{-\epsilon} \over 2D(D+\lambda)^3\epsilon} \frac{1}{16\pi^2},\nonumber \\
Z_{g_1}&=& 1+ {3g_1g_2\mu^{-\epsilon} \over 4\epsilon}\frac{1}{16\pi^2}-{D_0\lambda_1^2(5D+4\lambda)
\mu^{-\epsilon} \over D\lambda(D+\lambda)^2\epsilon}\frac{1}{16\pi^2}-{2DD_0\lambda_1^2
\mu^{-\epsilon} \over \lambda(D+\lambda)^3\epsilon}\frac{1}{16\pi^2}+{\lambda_1\chi g_1
(9D^2+16D\lambda +5\lambda^2)\mu^{-\epsilon} \over 2D(D+\lambda)^3\epsilon}\frac{1}{16\pi^2}.
 \label{fullz}\eea

There are no one-loop corrections to $\chi$ and
$\lambda$~\cite{comment}. To find out the $Z$-factor corresponding
to $\lambda_1$, we first set $\chi={D^2g_2\alpha \over \lambda_1}$
without any loss of generality, where $\alpha$ is a dimensionless
number.  From the relation $\Gamma^R_{\hat v\phi}=iZ_{\hat
v}Z_{\phi}k_i\chi=ik_i{D_R^2g_{2R} \alpha_R \over\lambda_{1R}}=i\chi
k_iZ_D^{-2}Z_{g_2}^{-1}Z_{\lambda_1}Z_{\alpha}^{-1}$, we find
$Z_{\hat v}=Z_D^{-2}Z_{g_2}^{-1}Z_{\lambda_1}Z_{\alpha}^{-1}
Z_{\hat\phi}^{-1}$. Now from $\Gamma^R_{\hat v_iv_i}({\bf k}=0)=
i\omega Z_vZ_{\hat v}=i\omega$ it can be easily seen that
$Z_vZ_{\hat v}=1$ or $Z_v=Z_{\hat v}^{-1}$. This lets us write
$Z_{\hat v}=Z_v^{-1}=Z_D^{-2} Z_{g_2}^{-1}Z_{\lambda_1}Z_\alpha
Z_{\hat\phi}^{-1}$. Next, using this relation in the expression
$\Gamma^R_{v_i\hat\phi\phi}=-ik_i\lambda_1 Z_{\lambda_1}^{-1}$, we
get \bea Z_\alpha &=& 1-{3g_1g_2\mu^{-\epsilon} \over
8\epsilon}\frac{1}{16\pi^2}-{\lambda_1g_1\chi (6D^2+8\lambda
D+3\lambda^2)\mu^{-\epsilon} \over
D(D+\lambda)^3\epsilon}\frac{1}{16\pi^2} +
{2\lambda_1^2D_0(2D^2+2\lambda^2+5D\lambda) \over \lambda
D(D+\lambda)^3\epsilon}\frac{1}{16\pi^2}\nonumber \\
&+&{\lambda_1g_1\chi(5D+3\lambda)\mu^{-\epsilon} \over
4D(D+\lambda)^2\epsilon}\frac{1}{16\pi^2} -
{\lambda_1^2g_1\chi^2(2D+\lambda)\mu^{-\epsilon} \over D^2\lambda
g_2
(D+\lambda)^2\epsilon}\frac{1}{16\pi^2}+{4\lambda_1^3D_0\chi\mu^{-\epsilon}
\over
\lambda^2D^2g_2(D+\lambda)\epsilon}\frac{1}{16\pi^2}.\label{zalpha}
\eea
 Equation~(\ref{zalpha}), together with $Z_v=Z_{\hat v}=1$ yield \bea
Z_{\lambda_1}=1+{g_1g_2\mu^{-\epsilon} \over
4\epsilon}\frac{1}{16\pi^2}+{\lambda_1g_1\chi
(3D+\lambda)\mu^{-\epsilon} \over
D(D+\lambda)^2\epsilon}\frac{1}{16\pi^2}+{\lambda_1g_1\chi
\mu^{-\epsilon} \over 2D(D+\lambda)\epsilon}\frac{1}{16\pi^2}.
\label{zlambda1}\eea
 Further, define $u=g_1g_2$ and $w={\lambda_1^2D_0 \over D^3}$, the $Z$
factors for them being $Z_u=Z_{g_1}Z_{g_2}$ and
$Z_w={Z_{\lambda_1}^2 \over Z_D^3}$; with $\theta={\lambda \over
D}$,  $Z_\theta=Z_D^{-1}$ as $\lambda$ does not renormalise in the
model. Formally, $Z$-factors (\ref{fullz},\ref{zalpha}) and
(\ref{zlambda1}) may be used to define the $\beta$-functions for the
renormalised coupling constants $u_R, w_R, \alpha_R, \theta_R$. We
obtain (after absorbing $1/16\pi^2$ in the definitions of the
renormalised coupling constants)
\begin{eqnarray}
&&\beta_u=u_R[\frac{3u_R}{2}-\frac{2w_R(5+4\theta_R)}{\theta_R
(1+\theta_R)^2} -
\frac{4w_R}{\theta_R(1+\theta_R)^3}+\frac{\alpha_R^2 u_R
(2+\theta_R)}{\theta_R (1+\theta_R)^2} -
\frac{4w_R\alpha_R}{\theta_R^2 (1+\theta_R)} + \frac{\alpha_R
u_R(3+\theta_R)}{(1+\theta_R)^2}\nonumber
\\&& + \frac{\alpha_R
u_R}{(1+\theta_R)^3}(7+14\theta_R+5\theta_R^2)-\epsilon],\label{betaufull}\\
&&\beta_w=w_R[\frac{7u_R}{8}+\frac{\alpha_Ru_R(7+3\theta_R)}{(1+\theta_R)^2}
-\frac{3\alpha_R u_R}{4(1+\theta_R)^3}(7+4\theta_R+\theta_R^2) -
\frac{6w_R}{\theta_R (1+\theta_R)^3}-\epsilon],\label{betawfull}\\
&&\beta_\alpha=\alpha_R[-\frac{3u_R}{8}-\frac{\alpha_R
u_R}{(1+\theta_R)^3}(6+8\theta_R +3\theta_R^2)+
\frac{2w_R(2+2\theta_R^2 + 5\theta_R)}{\theta_R
(1+\theta_R)^3}+\frac{\alpha_R u_R (5+3\theta_R)}{4(1+\theta_R)^3} -
\frac{\alpha_R^2 u_R (2+\theta_R)}{\theta_R (1+\theta_R)^2}\nonumber
\\&& +
\frac{4\alpha_Rw_R}{\theta_R^2 (1+\theta_R)}],\label{betaafull}\\
&&\beta_\theta=\theta_R[\frac{u_R}{8} - \frac{\alpha_Ru_R
(7+4\theta_R +\theta_R^2)}{4(1+\theta_R)^3} - \frac{2w_R}{\theta_R
(1+\theta_R)^3}]\label{betathetafull}.
\end{eqnarray}
The zeros of the $\beta$-functions
(\ref{betaufull}-\ref{betathetafull}) above should yield the fixed
points (FPs).
 Physically, there are three possible FP values for
$\theta_R$: $\theta_R=0,\infty$ and $\theta_R$ finite. The first two
should yield $z_\phi\neq z_v$ (weak dynamic scaling) and the last
one $z_\phi=z_v$ (strong dynamic scaling). In principle, FPs for all
the three physical regimes may be obtained from solutions of the
respective $\beta$-functions. However, due to the complicated natute
of Eqs.~(\ref{betaufull}-\ref{betathetafull}), the ensuing algebra
is rather involved, precluding full general solutions in closed
forms. Notice that that the $Z$-factors (\ref{fullz}, \ref{zalpha},
\ref{zlambda1}) simplify considerably in the limit
$\theta\rightarrow 0$ as shown below, corresponding to
$\theta_R\rightarrow 0$ in the renormalised theory. Instead of
obtaining the FPs with arbitrary values of $\theta_R$, we obtain the
FPs in the limit $\theta_R\rightarrow 0$ only.

\subsection{Analysis in the limit $\theta\rightarrow 0$}

By definition $\theta=\lambda/D$ or in terms of the renormalised
quantities $\theta_R=\lambda_R/D_R$. Thus, for $\theta_R\rightarrow
0$, $D_R\gg \lambda_R\Rightarrow z_\phi < z_v$, i.e., the dynamic
exponents of the two interacting field $\phi$ and $\bf v$ are
unequal. Thus, this corresponds to {\em weak dynamic scaling}.
Physically meaningful stable solution (corresponding to a stable FP)
should then reveal weak dynamic scaling with $z_\phi < z_v$. In
order to obtain the FPs for $\theta_R=0$, we consider the forms for
the $Z$-factors (\ref{fullz}, \ref{zalpha}, \ref{zlambda1}) in the
limit $\theta\rightarrow 0$ (or, equivalently, the limit
$\theta_R\rightarrow 0$ in the renormalised theory). The $Z$-factors
na\"ively reduce to

\begin{eqnarray}
Z_u&=&1+\frac{3u\mu^{-\epsilon}}{2\epsilon}\frac{1}{16\pi^2}-\frac{14
w\mu^{-\epsilon}}{\theta_R\epsilon}\frac{1}{16\pi^2}+
\frac{10u\alpha\mu^{-\epsilon}}{\epsilon}\frac{1}{16\pi^2} -
\frac{4w\alpha\mu^{-\epsilon}}{\theta^2\epsilon}\frac{1}{16\pi^2} +
\frac{2u\alpha^2\mu^{-\epsilon}}
{\theta\epsilon}\frac{1}{16\pi^2},\nonumber\\
Z_w&=& 1 +
\frac{7u\mu^{-\epsilon}}{8\epsilon}\frac{1}{16\pi^2}+\frac{7u\alpha\mu^{-\epsilon}}{4\epsilon}\frac{1}{16\pi^2}
-
\frac{6w\mu^{-\epsilon}}{\theta\epsilon}\frac{1}{16\pi^2},\nonumber \\
Z_\theta&=&1+\frac{u\mu^{-\epsilon}}{8\epsilon}\frac{1}{16\pi^2} -
\frac{7u\alpha\mu^{-\epsilon}}{4\epsilon}\frac{1}{16\pi^2}
- \frac{2w\mu^{-\epsilon}}{\theta\epsilon}\frac{1}{16\pi^2},\nonumber \\
Z_\alpha&=&1-\frac{3u\mu^{-\epsilon}}{8\epsilon}{1 \over 16\pi^2}-\frac{19u\mu^{-\epsilon}\alpha}{4\epsilon}
\frac{1}{16\pi^2}
+\frac{4w\mu^{-\epsilon}}{\theta\epsilon}{1 \over 16\pi^2}-
\frac{2u\mu^{-\epsilon}\alpha^2}{\theta\epsilon}\frac{1}{16\pi^2}
+
\frac{4w\mu^{-\epsilon}\alpha}{\theta^2\epsilon}\frac{1}{16\pi^2},\nonumber
\\
Z_{\lambda_1}&=&=1+\frac{u\mu^{-\epsilon}}{4\epsilon}\frac{1}{16\pi^2}+\frac{3u\alpha\mu^{-\epsilon}
}{\epsilon}\frac{1}{16\pi^2}+\frac{u\alpha\mu^{-\epsilon}}{2\epsilon}\frac{1}{16\pi^2}.\label{zsmalltheta}
\end{eqnarray}
The corresponding $\beta$-functions for the renormalised coupling
constants $u_R,w_R,\theta_R$ and $\alpha_R$ reduce to
\begin{eqnarray}
\beta_u&=&u_R[-\epsilon + \frac{3}{2} u_R-\frac{14 w_R}{\theta_R} +
10 u_R\alpha_R - 4\frac{w_R\alpha_R}{\theta_R^2} +
2\frac{u_R\alpha_R^2}{\theta_R}],\label{smallu}\\
\beta_w&=&w_R[-\epsilon+\frac{7u_R}{8}+\frac{7}{4}u_R\alpha_R -
\frac{6w_R}{\theta_R}],\label{smallw}\\
\beta_\alpha&=&\alpha_R[-\frac{3u_R}{8} -{19u_R\alpha_R \over 4} +
\frac{4w_R}{\theta_R}-\frac{2u_R\alpha_R^2}{\theta_R} +
\frac{4w_R\alpha_R}{\theta_R^2}],\label{smalla}\\
\beta_\theta&=&\theta_R[\frac{u_R}{8}-\frac{7}{4}u_R\alpha_R
-\frac{2w_R}{\theta_R}].\label{smalltheta}
\end{eqnarray}
 Since we consider the FPs in the $\theta_R\rightarrow 0$ limit,
 these  FPs are given by the zeros of the $\beta$-functions
 (\ref{smallu}-\ref{smalla}). The non-trivial FPs, for which $u_R,
 w_R, \alpha_R\neq 0$ are then given by the equations
\begin{eqnarray}
&&\frac{3}{2} u_R-\frac{14 w_R}{\theta_R} + 10 u_R\alpha_R -
4\frac{w_R\alpha_R}{\theta_R^2} +
2\frac{u_R\alpha_R^2}{\theta_R}=\epsilon,\label{nfpeq1}\\
&&\frac{7u_R}{8}+\frac{7}{4}u_R\alpha_R -
\frac{6w_R}{\theta_R}=\epsilon,\label{nfpeq2}\\
&&-\frac{3u_R}{8} -{19u_R\alpha_R \over 4} +
\frac{4w_R}{\theta_R}-\frac{2u_R\alpha_R^2}{\theta_R} +
\frac{4w_R\alpha_R}{\theta_R^2}=0.\label{nfpeq3}
\end{eqnarray}
Now, $u_R,\,w_R\sim O(\epsilon)$ and $\alpha_R$ is a finite number
(or zero) at the FPs with $\theta_R\rightarrow 0$. For physically
meaningful FPs with finite values of $u_R, w_R, \alpha_R$ in the
limit $\theta_R\rightarrow 0$, the $\beta$-functions
(\ref{smallu}-\ref{smalla}) should stay finite, or, equivalently, no
terms in the equations (\ref{nfpeq1}-\ref{nfpeq3}) should diverge.
Since $\theta_R$ appears in the denominators of several terms in the
$\beta$-functions (\ref{smallu}-\ref{smalla}), na\"ively those terms
diverge for $\theta_R\rightarrow 0$. To prevent this, the respective
numerators must scale with $\theta_R$ appropriately in the limit
$\theta_R\rightarrow 0$, so that the divergences in the various
terms of the $\beta$-functions (\ref{smallu}-\ref{smalla}) [or, in
Eqs.~(\ref{nfpeq1}-\ref{nfpeq3})] cancel out, and all terms in
(\ref{smallu}-\ref{smalla}) or in (\ref{nfpeq1}-\ref{nfpeq3}) are
finite (or zero) in the limit $\theta_R\rightarrow 0$.

To proceed further, we assume $w_R\sim\theta_R^{\gamma_1}$ and
$\alpha_R\sim\theta_R^{\gamma_2}$ in the limit $\theta_R\rightarrow
0$, with $\gamma_1,\gamma_2>0$ to be chosen such that the
divergences mentioned above are cancelled. Clearly, if
$\gamma_1,\gamma_2$ are too small, some of the terms in
Eqs.~(\ref{smallu}-\ref{smalla}) will still diverge for
$\theta_R=0$. On the other hand, if $\gamma_1,\gamma_2$ are too
large, then {\em all} terms with $w_R$ or $\alpha_R$ will vanish for
$\theta_R\rightarrow 0$, allowing only the DP FP to survive. 
Evidently, for finiteness of the $\beta$-functions in the limit
$\theta_R\rightarrow 0$ (\ref{smallu}-\ref{smalla}), we must have
$\gamma_1\geq 1,\,\gamma_1+\gamma_2 \geq 2,\,2\gamma_2 \geq 1$.
Non-trivial (non-DP) FPs are obtained, provided one or more of the above
inequalities reduce to equalities (i.e., the above conditions hold
with the "=" sign, instead of the "$\geq$" sign). Clearly, all of
them cannot hold good simultaneously with the "=" sign. Assuming any
two of the above three conditions should hold with "=" sign, i.e.,
any two of
$w_R/\theta_R,\,w_R\alpha_R/\theta_R^2,\,u_R\alpha_R^2/\theta_R$ are
not to vanish in the $\theta_R\rightarrow 0$ there are only two
sets of choices for $\gamma_1,\gamma_2$, for which non-trivial
(non-DP) FPs ensue, while keeping all the $\beta$-functions
(\ref{smallu}-\ref{smalla}) above finite. Noting that $u_R\sim
O(\epsilon)$ at the FP, the two choices are as follows:
\begin{itemize}
\item Case I: $\alpha_R\sim \sqrt\theta_R,\,w_R\sim \theta_R^{3/2}
\epsilon$. With this choice, $\alpha_R^2 u_R/\theta_R \sim
O(\epsilon)$ and $w_R/\theta_R\sim \sqrt\theta_R\epsilon\rightarrow
0$ for $\theta_R\rightarrow 0$. This corresponds to the bare
coupling constants $w\sim \theta^{3/2}\epsilon,\,\alpha\sim
\sqrt\theta$ for $\theta\rightarrow 0$.
\item Case II: $w_R\sim\theta_R\epsilon,\,\alpha_R\sim\theta_R$. Thus,
$w_R/\theta_R\sim O(\epsilon),\,\alpha_Rw_R/\theta_R^2 \sim
O(\epsilon)$ and $\alpha_R^2 u_R/\theta_R\sim
\theta_R\epsilon\rightarrow 0$ for $\theta_R\rightarrow 0$. This
corresponds to the bare coupling constants $w\sim
\theta\epsilon,\,\alpha\sim\theta$ when $\theta\rightarrow 0$.
\end{itemize}
We obtain the FPs separately for the two cases above. (Notice
that choices for $\gamma_1,\gamma_2$ such that only one among
$w_R/\theta_R,\,w_R\alpha_R/\theta_R^2,\,u_R\alpha_R^2/\theta_R$ is
not to vanish in the $\theta_R\rightarrow 0$ does not lead to any
new non-trivial FPs that are already not contained in Case I and
Case II above.)

Case I: Evidently, the $\beta$-functions (\ref{smallu}-\ref{smalla})
are linear in $u_R,u_R\alpha_R^2/\theta_R,w_R\alpha_R/\theta^2$.
Equivalently the $Z$-factors in (\ref{zsmalltheta}) are linear in
$u,u\alpha^2/\theta,w\alpha/\theta^2$. This allows us to identify
three effective (bare) coupling constants: (i) $u$,
(ii)$s=u\alpha^2/\theta$ (iii)$b=w\alpha/\theta^2$.  The
renormalisation $Z$-factors for $s$ and $b$ may be calculated in
straightforward ways. We obtain
\begin{eqnarray}
Z_s&=&Z_uZ_\alpha^2
Z_\theta^{-1}=1+\frac{5u\mu^{-\epsilon}}{8\epsilon}\frac{1}{16\pi^2}
+ \frac{4b\mu^{-\epsilon}}{\epsilon}\frac{1}{16\pi^2} -
\frac{2s\mu^{-\epsilon}}{\epsilon}\frac{1}{16\pi^2},\nonumber \\
Z_b&=&Z_wZ_\alpha
Z_\theta^{-2}=1+\frac{u\mu^{-\epsilon}}{4\epsilon}\frac{1}{16\pi^2}
+ \frac{4b\mu^{-\epsilon}}{\epsilon}\frac{1}{16\pi^2} -
\frac{2s\mu^{-\epsilon}}{\epsilon}\frac{1}{16\pi^2},\nonumber \\
Z_u&=&1+\frac{3u\mu^{-\epsilon}}{2\epsilon}\frac{1}{16\pi^2}-
\frac{4b\mu^{-\epsilon}}{\epsilon}\frac{1}{16\pi^2}+\frac{2s\mu^{-\epsilon}}{\epsilon}\frac{1}{16\pi^2}.
\end{eqnarray}
The corresponding $\beta$-functions for the remormalised coupling
constants $u_R,b_R,s_R$ are given by (again absorbing $1/16\pi^2$)
\begin{eqnarray}
\beta_u&=&u_R[-\epsilon + \frac{3u_R}{2}-4b_R +
{2s_R }],\label{betaunew}\\
\beta_s&=&s_R[-\epsilon + \frac{5u_R}{8} + 4b_R -
{2s_R}],\label{betanews}\\
\beta_b&=&b_R[-\epsilon + \frac{u_R}{4}+4b_R -
{2s_R}].\label{betanewb}
\end{eqnarray}
The FPs, as usual, are given by the zeros of the $\beta$-functions
(\ref{betaunew}-\ref{betanewb}). Notice that $s_R$ and $u_R$ must
have the same sign and $b_R$ can be of any sign. The FPs are given
by
\begin{itemize}
\item FPI: Gaussian FP - $u_R=0,\,s_R=0,\,b_R=0$.
\item FPII: DP FP - $u_R=\frac{2\epsilon}{3},\,s_R=0,\,b_R=0$.
\item FPIII: $u_R=0,\,s_R=0,\,b_R=\epsilon/4$.
\item FPIV: $s_R=0,\,u_R=8\epsilon/7,\,b_R=5\epsilon/28$.
\item FPV: $b_R=0,\,u_R={16\epsilon \over 17},\,s_R=-{7\epsilon \over 34}$. This is
unphysical, since $s_R$ and $u_R$ have different signs. Therefore,
we discard this.
\end{itemize}
We now analyse the stability of the above FPs by finding the
eigenvalues  $\Lambda$ of the stability matrix corresponding to each
physically meaningful FP. We find
\begin{itemize}
\item FPI (Gaussian FP): The eigenvalues are
$\Lambda=-\epsilon,-\epsilon,-\epsilon$. The negativity of the all
the eigenvalues indicate that this FP is unstable in all directions.
\item FPII (DP FP): The eigenvalues are $\Lambda=\epsilon, {-5\epsilon
\over 6},{-7\epsilon \over 12}$. Thus, it  is only stable along the
$u_R$-axis and unstable in the other directions at this coupling
constant space.
\item FPIII The eigenvalues are $\Lambda=-2\epsilon,\epsilon,0$.
Thus, FPIII is unstable along the $u_R$-direction.
\item FPIV: We find that this is stable along the $s_R$-axis and stable and
oscillating along the $u_R-b_R$ plane in the space of renormalised
coupling constants $u_R,\,b_R,\,s_R$: the eigenvalues $\Lambda$ for
the corresponding stability matrix are given by
\begin{equation}
\Lambda={17\epsilon \over 14}+i{5.56\epsilon \over 14},{17\epsilon
\over 14} -i{5.56\epsilon \over 14},{3\epsilon \over 7},
\end{equation}
showing positivity of the real parts of the eigenvalues (hence
stable). This leaves us with FPIV as the only stable FP for Case I.
\end{itemize}
 We now obtain the corresponding critical exponents.
To find out the critical exponents corresponding to these fixed
points we need to evaluate the Wilson's flow functions which are
defined as \bea \zeta_\phi=\mu{\partial \over
\partial\mu}\ln{Z_\phi}\,,\, \zeta_{\hat\phi}= \mu{\partial \over
\partial\mu}\ln{Z_{\hat\phi}}\,,\, \zeta_D=\mu{\partial \over
\partial\mu} \ln{Z_D}\,,\, \zeta_\tau=\mu{\partial \over
\partial\mu}\ln{Z_\tau}-2.\label{flow} \eea
 From the flow functions in Eq.~(\ref{flow}), the critical exponents
of the model in terms of the renormalised coupling constants can be
obtained as shown below. \bea
\eta_\phi &=& \eta_{\hat\phi}=-\zeta_\phi \\
{1 \over \nu} &=& -\zeta_\tau \\
z_\phi &=& 2-\zeta_D. \eea We find
\begin{itemize}
\item FPII (DP FP): $\eta_\phi=\eta_{\hat\phi}={\epsilon \over 12}$, $\nu^{-1}=2+{\epsilon \over 4}$,
dynamic exponent $z_\phi=2-{\epsilon \over 12}$.
\item FPIII: Dynamic exponent $z_\phi = 2-\zeta_D=2$. Evidently, this is in
contradiction with the expected weak dynamic scaling for
$\theta_R\rightarrow 0$ (see discussions above).
\item FPIV: $z_\phi=2-\zeta_D=2-u_R/8=2-\epsilon/7 <2$. As we shall see below, for $z_\phi <2$, $z_v=2$, which
implies weak dynamic scaling, consistent with $\theta_R \rightarrow
0$. Thus, this FP represents {\em weak dynamic scaling}. Other
critical exponents are (a) anomalous dimension
$\eta_\phi=-\zeta_\phi=u_R/8=\epsilon/7$, (b) inverse correlation
length exponent $1/\nu= -\zeta_\tau = 3 u_R/8 + 2= 3\epsilon/7 +2$.
\end{itemize}
Thus, FPIV is a stable FP in the coupling constant space spanned by
$u_R,s_R,b_R$, that describes weak dynamic scaling at the AAPT, in
accordance with our stating assumption $\theta_R\rightarrow 0$. In
addition, note that $\partial\beta_\theta/\partial\theta_R=u_R/8>0$
and $\partial\beta_\theta/\partial {\mathcal Y}|_{\rm FPIV}=0$,
where ${\mathcal Y}=u_R,b_R,s_R$. This indicates that the weak
dynamic scaling behaviour represented by FPIV is indeed stable along
the $\theta_R$ direction as well. Lastly, at the DP FP (FPII), all
the critical exponents are unsurprisingly identical to their values
for the usual DP problem. Since FPII is an unstable FP in the
present model, due to the coupling of $\phi$ with the environment
(modeled by $\bf v$), the DP FP of the original DP problem gives way
to a non-DP FP that characterises the underlying AAPT.

Consider now Case II: With the scaling of $w_R$ and $\alpha_R$ with
$\theta_R$ for Case II, evidently the $\beta$-functions
(\ref{smallu}-\ref{smalla}) are linear in the {\em effective
coupling constants} $u_R,\,m_R=w_R/\theta_R,\,\psi_R=\alpha_R
w_R/\theta_R^2$. The corresponding $\beta$-functions are obtained in
a straightforward way: $\beta_a=\mu\frac{\partial}{\partial\mu}
a_R,\,a=u,m,\psi$. These $\beta$-functions can be used to find out
the FPs present in the model by setting their values to zero. We
obtain \bea
Z_m &=& Z_wZ_\theta^{-1}=1+{3u\mu^{-\epsilon} \over 4\epsilon}\frac{1}{16\pi^2}-{4m\mu^{-\epsilon} \over \epsilon}\frac{1}{16\pi^2}, \\
Z_\psi &=& Z_\alpha Z_wZ_\theta^{-2}=1+{u\mu^{-\epsilon} \over
4\epsilon}\frac{1}{16\pi^2}+{2m\mu^{-\epsilon} \over
\epsilon}{1 \over 16\pi^2}+{4\psi\mu^{-\epsilon} \over \epsilon}\frac{1}{16\pi^2}.
\eea From these $Z$-factors effective $\beta$-functions for the
corresponding renormalised coupling constants can be written down
easily given by \bea
\beta_u &=& u_R\left[-\epsilon +{3u_R \over 2}-14m_R-4\psi_R\right],\label{betau1} \\
\beta_m &=& m_R\left[-\epsilon +{3u_R \over 4}-4m_R\right], \label{betaphi1}\\
\beta_\psi &=& \psi_R\left[-\epsilon +{u_R \over 4}+2m_R +4\psi_R\right], \label{betapsi1}
%\beta_\theta &=& \theta_R\left[{u_R \over 8}-2\phi_R\right]
\eea where factors of $1/16\pi^2$ have been absorbed in the
definitions of the renormalised coupling constants. The zeros of the
$\beta$-functions (\ref{betau1}-\ref{betapsi1}) yield a number of
FPs. We give the details below. Excluding the Gaussian FP, we have
\begin{itemize}
\item FPV: Set $m_R=\psi_R=0,u_R\neq0$, giving
us the usual DP FP - $u_R={2\epsilon \over 3}$.

\item FPVI: Consider  $u_R=0,m_R\neq 0,\psi_R\neq 0$.
We find $m_R=-{\epsilon \over 4},\psi_R={3\epsilon \over 8}$. Now,
$m=w/\theta$, $m_R=w_R/\theta_R$ cannot be negative, since both
$w_R,\theta_R$ are non-negative. Thus, this FP is unphysical.

\item FPVII: Now consider $u_R\neq 0,m_R=0,\psi_R\neq 0$. The fixed points obtained
are $u_R={8\epsilon \over 7}, m_R=0, \psi_R={5\epsilon \over 28}$.

\item FPVIII: Next consider $\psi_R=0,u_R\neq 0,m_R\neq 0$. We obtain
$u_R={20\epsilon \over 9}, m_R={\epsilon \over 6}$.
\item FPIX: Lastly, we obtain another FP
$u_R=2\epsilon$, $m_R={\epsilon \over 8}$ and $\psi_R={\epsilon
\over 16}$, when all the effective coupling constants are non-zero at the FP.
\end{itemize}

Having derived all the relevant FPs from the $\beta$-functions, we
analyse the stability of these FPs by finding the eigenvalues of the
stability matrix corresponding to each physically meaningful FP. We
find
\begin{itemize}
\item FPV (DP FP): the eigenvalues of the stability matrix are
$\Lambda=\epsilon, -{\epsilon \over 2},-{5\epsilon \over 6}$. The
positivity of the eigenvalue along the $u_R$-direction indicates
that it is stable along the $u_R$ axis but it is unstable along the
$m_R$ and the $\psi_R$ axes.
\item FPVII: the eigenvalues of the stability matrix are $\Lambda={10\epsilon \over 7}, \epsilon$
along the $u_R-\psi_R$ plane and $\Lambda=-{\epsilon \over 7}$ along the $m_R$ axis.
This shows that the FP is stable along the $u_R-\psi_R$ plane but unstable as expected
along the $m_R$ direction.
\item FPVIII: the
eigenvalues of the stability matrix are $\Lambda=\epsilon,{5\epsilon \over 3},
-{\epsilon \over 9}$. This shows that
FP is stable in the $u_R-m_R$ plane but are unstable along the
$\psi$ axis.
\item FPIX:, the
eigenvalue equation yields
$\Lambda=1.5752\epsilon,0.059\epsilon,1.115\epsilon$. As all the
eigenvalues are positive, this FP is stable in the whole
$u_R-m_R-\psi_R$ parameter space.
\end{itemize}
Therefore, we find that the nontrivial FP characterised by non-zero
$u_R,m_R,\psi_R$ is stable in {\em all} three directions in the
space of the three coupling constants. Nonzero $m_R$ and $\psi_R$ at
the FP suggest nonzero $w_R$ and $\alpha_R$ at the FP, indicating
their relevance in a DRG sense. Thus, overall, both the environment
and the feedback on it are relevant in determining the macroscopic
scaling at the AAPT. Furthermore, our analyses above are limited
only to the case $\theta\rightarrow 0$. Notice that with the
obtained values $u_R=2\epsilon,m_R=\epsilon/8,\psi_R=\epsilon/16$ at
the nontrivial FP, $\partial\beta_\theta/\partial\theta=0$, making
$\theta_R=0$ {\em marginal} at the FP. Thus, we are unable to
comment whether $\theta_R=0$ is a stable FP or not, although its
instability cannot be ruled out on any general ground.

To find out the critical exponents corresponding to these fixed
points we need to evaluate the Wilson's flow functions as defined in
Eqs.~(\ref{flow}) above. Using the definitions of the critical
exponents in terms of the flow functions,  we obtain
\begin{itemize}
 \item FPV or the DP FP (${2\epsilon \over 3},0,0$): \\
$\eta_\phi=\eta_{\hat\phi}={\epsilon \over 12}$, $\nu^{-1}=2+{\epsilon \over 4}$,
$z_\phi=2-{\epsilon \over 12}$.
\item FPVII (${8 \epsilon \over 7},0,{5\epsilon \over 28}$): \\
$\eta_\phi=\eta_{\hat\phi}={\epsilon \over 7}$, $\nu^{-1}=2+{3\epsilon \over 7}$,
$z_\phi=2-{\epsilon \over 7}$.
\item  FPVIII (${20\epsilon \over 9},{\epsilon \over 6},0$): \\
$\eta_\phi=\eta_{\hat\phi}={\epsilon \over 9}$,
$\nu^{-1}=2+{\epsilon \over 2}$, $z_\phi=2+{\epsilon \over 18}$.
Thus, $z_\phi > 2$. %This is in contradiction with our starting
%assumption of $\theta\rightarrow 0$. Hence, we discard this FP.
\item  FPIX ($2\epsilon,{\epsilon \over 8},{\epsilon \over 16}$): \\
$\eta_\phi=\eta_{\hat\phi}={\epsilon \over 8}$, $\nu^{-1}=2+{\epsilon \over 2}$,
$z_\phi=2$.
\end{itemize}
Notice that at FPs, FPVIII and FPIX,
where the effects of the environment and the feedback on it are
relevant in a DRG sense, $z_\phi>2$ and $z_\phi=2$, respectively. As
we see in the next Section, for $z_\phi\geq 2$, the dynamic exponent
for $v_i$, $z_v=z_\phi$, indicating strong dynamic scaling at these
FPs. This is in contradiction with the expected weak dynamic scaling
at $\theta_R=0$; in other words $z_\phi<z_v$ is expected. Therefore,
FPVIII and FPIX are unphysical FPs. In contrast at FPVII,
$z_\phi <2$; with $z_v=2$ (see below) this corresponds to weak
dynamic scaling. However, it is unstable along the $m_R$-direction. 
Hence, we find that there is only one FP (FPIV) that is stable in
all directions and describe weak dynamic scaling for the AAPT, and
so represents a physically correct scaling behaviour at the AAPT.

\subsection{Scaling exponents of the broken symmetry
field}\label{scalev}

To obtain the scaling exponents of the broken symmetry field $\bf
v$, we start from Eq.~(\ref{equ}) for $v_i$. Evidently, if $\chi=0$,
i.e., if the dynamics of $v_i$ is autonomous, $v_i$ can be solved
exactly with
 \beq \langle v_i ({\bf k},\omega)v_j ({\bf -k},-\omega)\rangle = \frac{2D_0\delta_{ij}}{\omega^2
 + \lambda^2 k^4},\,\langle v_i ({\bf k},t) v_j ({\bf -k},t)\rangle =
 \frac{D_0}{\lambda k^2},
 \eeq
 and
hence the exponents of $v_i$ are also known exactly: Dynamic
exponent $z_v=2$ and anomalous dimension $\eta_v=0$. When $\chi\neq
0$, one can still obtain an exact closed form for the correlator of
$v_i$, owing to the linearity of the feedback term in
Eq.~(\ref{equ}): \bea
 \langle v_i ({\bf k},\omega)v_j ({\bf -k},-\omega)\rangle =
 \frac{2D_0\delta_{ij}}{\omega^2 + \lambda^2 k^4} +
 \frac{\chi^2 \langle |\phi({\bf k},\omega)|^2\rangle k_i
 k_j}{\omega^2 + \lambda^2 k^4}.
 \eea
 Noting that in terms of the scaling exponents and in terms of the
 renormalised parameters
 \beq \langle |\phi ({\bf k},\omega)|^2\rangle \sim
 \frac{1}{k^{2-\eta_\phi}}\frac{D_R k^{z_\phi}}{\omega^2 + D_R^2
 k^{2z_\phi}} \eeq
 leading to
\beq \langle v_i ({\bf k},t) v_j ({\bf -k},0)\rangle \sim
\frac{\chi^2 k_i k_j D_R k^{z_\phi}}{k^{2-\eta_\phi}}\left[
\frac{\exp (-D_Rk^{z_\phi}t)}{2D_R k^{z_\phi} (\lambda^2 k^4 -D_R^2
k^{2z_\phi})} + \frac{\exp (-\lambda k^2 t)}{2\lambda k^2 (D_R^2
k^{2z_\phi} - \lambda^2 k^4)}\right] + \frac{D_0 \exp (-\lambda k^2
t)\delta_{ij}}{2\lambda k^2},\eeq yielding in the hydrodynamic limit
$k\rightarrow 0$ and, assuming $z_\phi<2$, for large time $t\gg
1/(Dk^{z_\phi})$,
 \beq \langle v_i ({\bf k},t)v_j ({\bf -k},0)\rangle \sim \frac{\exp
 (-\lambda k^2 t)\delta_{ij}}{2\lambda k^2} + \frac{\chi^2 k_i
 k_j D_R k^{z_\phi}}{k^{2-\eta_\phi}} \frac{\exp (-\lambda k^2
 t)}{2\lambda k^2 [D_R k^{2z_\phi} - \lambda^2 k^4]},\eeq
 giving a dynamic exponent $z_v =2$, and hence weak dynamic scaling.
  On the other hand for $z_\phi=2$, one, of course, has $z_v=2$,
 displaying strong dynamic scaling. Furthermore, in the event $z_\phi >2$, it is clear from the preceding discussion
 that $z_\phi=z_v$, indicating strong dynamic scaling. We can further obtain results
 for
 the anomalous dimension $\eta_v$ of $v_i$: We have for the
 equal-time correlator
 \beq \langle v_i ({\bf k},t)v_j ({\bf -k},t)\rangle \sim
 \frac{D_0\delta_{ij}}{\lambda k^2} + \frac{\chi^2 k_i
 k_j}{k^{2-\eta_\phi}}\frac{1}{D_R k^{z_\phi}}\label{vstat}.\eeq
 Thus, if $D_0=0$ then $\eta_v = \eta_\phi - z_\phi-2$. However, if
 $D_0\neq 0$, then if $ -\eta_\phi+z_\phi -2<0$ then the first term
 on the right hand side of (\ref{vstat}) dominates, giving
 $\eta_v=0$, else $\eta_v=\eta_\phi - z_\phi+2$. Thus, at the nontrivial FP (FPVIII),
 $\eta_v={2\epsilon \over 7}$. This completes the
 discussions on the enumeration of the scaling exponents of $\bf v$.

%\subsection{General case: $m_R\neq 0\neq w_R$}

\section{summary and outlook}
\label{summ}

In this article, we have constructed a simple model and studied it
to find  how the mutual interactions between a density undergoing an
AAPT and its surrounding fluctuating environment affect the
universal scaling properties of both the density field and the
environment at the extinction transition. We have used a broken
symmetry mode, modeled by a vector field, to represent the
fluctuating environment. We have used a perturbative (up to the
one-loop order) DRG calculation to extract the relevant scaling
exponents that define the AAPT. The zeros of the DRG
$\beta$-functions yield DRG fixed points, each representing a phase,
characterised by a set of values for the scaling exponents. Due to
the algebraic complications involved, we have been able to obtain
the FPs only in the limit $\theta_R\rightarrow 0$. This corresponds
to weak dynamic scaling with $z_\phi<z_v$. Thus, the possibility of
$z_\phi>z_v$ is effectively ignored. We obtain one FP (FPIV above)
with $\theta_R\rightarrow 0$ that yields physically acceptable
results for the scaling exponents at the AAPT. This FP is stable in
all the directions in the space of the effective coupling constants,
and also in the direction of $\theta_R$. Thus, we  speculate that
our model displays only weak dynamic scaling with the scaling
properties at the AAPT being given by FPIV. The quantitative
accuracy of the scaling exponents obtained are limited by the
approximations involved. Nonetheless, since the exponents at FPIV is
different from their usual DP counterparts, we are able to show that
both the environmental influence and the feedback on it by the
density undergoing the AAPT are generally relevant in a DRG sense.
While we speculate about our model displaying only weak dynamic
scaling, existence of stable FPs with finite $\theta_R$
corresponding to strong dynamic scaling (i.e., $z_\phi=z_v$) should
be investigated numerically from the zeros of the $\beta$-functions
(\ref{betaufull}-\ref{betathetafull}) with finite $\theta_R$ as
complementary to the present study. We briefly discuss the
possibility of FPs in Appendix~\ref{largetheta} with
$\theta_R\rightarrow\infty$ (i.e, with $z_\phi>z_v$). We show that
there are no stable FPs there. Regardless of the limitations of our
calculations here, generally at the physically acceptable stable FP
obtained here, at which the feedback is relevant, not only the
scaling exponents of the density field undergoing AAPT are affected
by the environment, even the scaling exponents of the coupled broken
symmetry mode (the environment) should be affected in turn. This
clearly establishes the relevance of feedback (in a DRG sense) for
both the density and the broken symmetry fields.

As discussed above, our model equation (\ref{equ}) and its symmetry
(polar symmetry) are simplified versions of realistic models. More
realistic models include, e.g., the equations of motion of polar
order parameter of an active (nonequilibrium) polar system, which
couples to the concentration of the active particles in a way
similar to Eq.~(\ref{equ}) above~\cite{toner}. However, the
structure of the polar order parameter equation in an active system
is much more complicated~\cite{toner}, than the simplified equation
for $\bf v$ that we have used here. It would be interesting to
investigate how different symmetries of $\bf v$ (e.g., polar versus
nematic) may change the emerging scaling behaviour at the critical
point. Our work highlights the importance of feedback of the density
undergoing AAPT on the environmental dynamics. However, our study
here is confined to illustrating the effects of linear feedback.
This would be relevant, e.g., in a bacteria colony in its ordered
state undergoing birth and death. There may, however, be situations
where the feedback is nonlinear. An interesting example could be the
AAPT of a density field being advected by an incompressible velocity
field~\cite{nelsonpap} or the birth-growth of bacteria in their
nematic ordered state.  It will be theoretically interesting to
study the effects of nonlinear feedback, especially in the context
of weak and strong dynamic scaling. Our work should also be useful
in understanding other realistic situations, e.g., extinction
transition in an orientationally ordered bacteria film resting on a
fluctuating surface or a fluctuating membrane.

\section{Acknowledgement}
One of the authors (AB) wishes to thank  the Max-Planck-Society
(Germany) and the Department of Science and Technology/Indo-German
Science and Technology Centre (India) for partial financial support
through the Partner Group programme (2009).

\appendix
\section{Large $\theta_R$ limit} \label{largetheta}

We here consider the stability of the DP FP in the limit
$\theta_R\rightarrow \infty$ (equivalently, bare
$\theta\rightarrow\infty$). In this limit, we should get
$z_\phi>z_v$ (again weak dynamic scaling, but the opposite of FPIV).
Let $\overline \theta =1/\theta$, so that
$\overline\theta\rightarrow 0$. In this limit, no divergences are
encountered in the $Z$-factors (\ref{fullz}-\ref{zalpha}), or, in
the $\beta$-functions (\ref{betaufull}-\ref{betathetafull}) and the
limit $\overline\theta\rightarrow 0$ may be taken directly and
smoothly. The relevant $\beta$-functions for the renormalised
coupling constants $u_R,\,w_R,\,\alpha_R$ in this limit are
\begin{eqnarray}
\beta_u&=&u_R[-\epsilon +\frac{3u_R}{2}],\\
\beta_w&=&w_R[-\epsilon +\frac{7u_R}{8}],\\
\beta_\alpha&=&\alpha_R[-\frac{3u_R}{8}],\\
\beta_{\overline\theta}&=&\overline\theta_R[-\frac{u_R}{8}].
%u_R\overline\theta_R}{4}].
\end{eqnarray}
Thus, apart from the trivial Gaussian FP, the only other FP is
$u_R=2\epsilon/3,w_R=0,\alpha_R=0$ (DP FP) together with $\overline
\theta_R=0$. It is not surprising that with $\overline\theta_R=0$,
DP FP is the only non-zero FP left in the system, since with
(assumed) $z_\phi>z_v$ $\bf v$-fluctuations vanish for time-scales
$t\gg 1/(\lambda k^{z_v})$. However, this FP is {\em unstable} in
all the three directions of $w_R,\alpha_R,\overline\theta_R$. Again,
this is consistent with our argument in Sec.~\ref{scalev} that
$z_\phi \leq z_v$ necessarily, precluding the possibility of
$z_\phi>z_v$, as it would be for $\overline\theta_R\rightarrow 0$.
Our analysis here however does not rule out the possibility of FPs
with non-zero but finite $\theta_R$ (i.e., with strong dynamic
scaling  $z_v=z_\phi$). Such FPs, if exist, may be analysed by
numerically solving for the zeros of the $\beta$-functions
(\ref{betaufull}-\ref{betathetafull}). We do not do this here.

%\section{appendix} give the feynman diagrams?


\begin{thebibliography}{99}
\bibitem{crystal}  Although this is similar to that for the
displacement field in a crystal, our purpose is not to specifically
model the environment in terms of a crystalline order.
\bibitem{dpcalc} Analogous transformations bring the action
functional corresponding to the pure DP problem to a form, where the
rapidity symmetry manifests easily~\cite{uwe-hans-review}. While
there is no rapidity symmetry in the present model (see below), we
nonetheless perform the above transformation for calculational
convenience and easy comparison with the corresponding calculation
for the DP process at the intermediate steps.
\end{thebibliography}
\end{document}